\documentclass{aa}  

\usepackage{graphicx}
\usepackage{txfonts}
\usepackage{lipsum}
\usepackage{subcaption}         
\usepackage{hyperref}
\usepackage[capitalize]{cleveref}
\usepackage{lscape}      
\usepackage{amsmath}
\usepackage{placeins}           
\usepackage{natbib}

\hypersetup{
        colorlinks=true,    
        linkcolor=blue,  
        citecolor=blue,    
        filecolor=cyan,     
        urlcolor=black,      
} 

\begin{document}

   \title{Cosmic Vine: High abundance of massive galaxies and dark matter halos in a forming cluster at z=3.44}
   \titlerunning{Massive galaxies and dark matter halos in a protocluster at $z=3.44$}
   \authorrunning{Sillassen et al.}
   
   \author{
   Nikolaj B. Sillassen\inst{1,2}\thanks{Corresponding authors: nbsi@space.dtu.dk, shuji@dtu.dk} 
   \and 
    Shuowen Jin\inst{1,2}\protect\footnotemark[1]
    \and 
    Georgios E. Magdis\inst{1,2}
    \and
    Francesco Valentino\inst{1,2}
    \and
    Emanuele Daddi\inst{3}
    \and
    Raphael Gobat\inst{4}
    \and
    Malte Brinch\inst{5,6}
    \and
    Kei Ito\inst{1,2}
    \and
    Tao Wang\inst{7,8}
    \and 
    Hanwen Sun\inst{7,8}
    \and
    Gabriel Brammer\inst{1,9}
    \and
    Sune Toft\inst{1,9}
    \and
    Thomas R. Greve\inst{1,2}
    }
   \institute{Cosmic Dawn Center (DAWN), Denmark
            \and DTU Space, Technical University of Denmark, Elektrovej 327, DK-
            2800 Kgs. Lyngby, Denmark
            \and
            CEA, IRFU, DAp, AIM, Université Paris-Saclay, Université Paris Cité, Sorbonne Paris Cité, CNRS, 91191 Gif-sur-Yvette, France
            \and
            Instituto de Física, Pontificia Universidad Católica de Valparaíso, Casilla 4059, Valparaíso, Chile
            \and
            Instituto de Física y Astronomía, Universidad de Valparaíso, Avda. Gran Bretaña 1111, Valparaíso, Chile
            \and
            Millennium Nucleus for Galaxies (MINGAL), Chile
            \and
            School of Astronomy and Space Science, Nanjing University, Nanjing 210093, China
            \and
            Key Laboratory of Modern Astronomy and Astrophysics, Nanjing University, Ministry of Education, Nanjing 210093, China
            \and
            Niels Bohr Institute, University of Copenhagen, Jagtvej 128, 2200 Copenhagen, Denmark
            }

   \date{Received xx, 20XX}
 
\abstract{The Cosmic Vine is a massive protocluster at $z=3.44$ in the JWST CEERS field, offering an ideal laboratory for studying the early phases of cluster formation. Using the data from the DAWN JWST Archive, we conducted a comprehensive study on the large-scale structure, stellar mass function (SMF), quiescent members, and dark matter halos in the Cosmic Vine.
First, we spectroscopically confirmed 136 galaxies in the Vine at $z\approx3.44$, along with an additional 47 galaxies belonging to a diffuse foreground structure at $z\approx3.34,$ which we dubbed the Leaf. 
We identified four subgroups comprising the Cosmic Vine and two subgroups within the Leaf.
Second, we identified 11 quiescent members with $\log(M_*/{\rm M_\odot})=9.5-11.0$, the largest sample of quiescent galaxies in overdense environments at $z>3$, which gives an enhanced quiescent galaxy number density $\sim1-2\times10^{-4}$~cMpc$^{-3}$ that is two to three times above the field level at $\log(M_*/{\rm M_\odot})>10$. Notably, these quiescent members form a tight red sequence on the colour-magnitude diagram, making it one of the earliest red sequences known to date.
Third, by constructing the SMFs for both star-forming and quiescent members, we find that both SMFs are top-heavy, with a significantly enhanced quiescent fraction at $\log(M_*/{\rm M_\odot})>10.5$ compared to field counterparts. The stellar mass–size analysis reveals that star-forming members are more compact at higher masses than their field counterparts. Finally, we estimated a halo mass of $\log(M_{\rm h}/{\rm M_\odot})=13.2\pm0.3$ for the protocluster core and $\log(M_{\rm h}/{\rm M_\odot})=11.9-12.4$ for satellite subgroups. The phase-space analysis indicates that three subgroups are likely infalling to the core.
This work reveals a high abundance of massive galaxies and dark matter halos in this forming cluster, demonstrating the accelerated assembly of massive galaxies in massive halos when the Universe was less than 2 billion years old.

 }

   \maketitle

\section{Introduction}
Clusters of galaxies are the most massive structures in the Universe bound by the immense gravity of a virialised dark matter halo.
The galaxy population in local galaxy clusters is dominated by old gigantic ellipticals that have scarce ongoing star formation, which tightly assemble in a 'red sequence' in the colour-magnitude space \citep{Kodama1998_red_sequence,Valentinuzzi2011,Wetzel2012_red_sequence}.
The question of how these massive cluster galaxies formed and when they ceased their star formation remain open questions, for which the key to providing answers lies in probing the progenitors of clusters at early cosmic time.
Protoclusters, representing the densest structures in the early Universe with accelerated galaxy formation and evolution, are promising progenitors of local galaxy clusters \citep{Chiang2013cluster,Overzier2016}. 
Massive galaxies in protoclusters hold critical clues for the formation of brightest cluster galaxies (BCGs) and massive ellipticals in local clusters, while massive quiescent members are particularly important for studying the emergence of the red sequence and testing the models of galaxy quenching and environmental effects \citep{Strazzullo2016,Chartab2020,Kubo2021quiescent,Ito2023,Ito2025,Afanasiev2023_carla_mass_size,Sun2024_top_heavy_cl1001,Sun2025_bigfoot,Xu2025_rps_in_cl1001}.
Cosmological simulations predict that protoclusters form within dense nodes of the filamentary cosmic web at early cosmic times and grow through accretion of cold gas and mergers of galaxies and dark matter halos \citep{Dekel2009Nature,Mandelker2020,Schaye2023,Montenegro_Taborda2023,Nelson2024}. 
Observations have revealed the large-scale cosmic web and extreme overdensities of galaxies at high redshift, showing a coherent general picture as seen in simulations. As yet, the models have not yet been thoroughly tested and simulations have been seen with clear shortcomings compared with observations \citep[e.g.][]{Shen2022,Kimmig2023,Jin2024_cosmic_vine}. For example, current simulations often fail to realistically reproduce protoclusters and massive galaxies within them, preventing us from investigating detailed physics of cluster galaxy formation and the environmental effects that shape their evolution \citep[e.g.][]{Granato2015simu,Bassini2020,Jin2024_cosmic_vine}. Dark matter halos play a fundamental role in structure formation, but assessing their masses and properties is extremely challenging in observations.
Therefore, deep observations in high-$z$ protocluster fields and improved halo mass estimates are crucial for unveiling both the bright and dark sides of early cluster formation. To date, massive protoclusters have been discovered at the epoch of cosmic noon ($z\sim2$) up to cosmic dawn ($z\sim6$) using various methods (e.g.  \citealt{Pentericci2000,Gobat2011,Capak2011Nature,Walter2012,Toshikawa2016,Wang_T2016cluster,Oteo2018cluster,Miller2018cluster_z4,Pavesi2018,McConachie_2022,Zhou2024,Jin2024_cosmic_vine,Sillassen2024,Shah2024}). 

However, due to the limited sensitivity and wavelength coverage of the \textit{Hubble} Space Telescope (HST) and ground-based facilities, our view of these protoclusters is based on only a small number of galaxies that could be confirmed with spectroscopy, such as Ly$\alpha$ emitters (LAEs), H$\alpha$ emitters \citep[HAEs; e.g.][]{Shimakawa2014HAE,Guaita2020,Ramakrishnan2023,Brinch2024,UrbanoStawinski2024,PerezMartinez2025}, and dusty star-forming galaxies (DSFGs; \citealt{Daddi2009GN20,Dannerbauer2014LABOCA,Oteo2018cluster,Miller2018cluster_z4,Gomez-Guijarro2019,Harikane2019,Jin2021cluster,Sillassen2022,Alberts2022}). 
Without highly complete memberships and deep multi-wavelength observations, it is not feasible to robustly constrain the global properties (e.g. total stellar masses),  subtle environmental effects, and dark matter halos. 
Previous studies have found no significant association between biased tracers (e.g. quasars) and the density field \citep[e.g.][]{Husband2013,Uchiyama2018}; thus, it is crucial to test this with deeper and more complete samples (e.g. \citealt{Eilers2024}).
 JWST has provided a great opportunity to discover protoclusters at high redshift and, in particular, massive structures have been discovered in JWST deep fields with highly complete memberships \citep[e.g.][]{Sun2024_top_heavy_cl1001,Jin2024_cosmic_vine,Sun2025_bigfoot}.

These discoveries have revealed large-scale structures that span tens of Mpc \citep{Jin2024_cosmic_vine,Li_2025_epochsx}. Massive quiescent galaxies have been discovered in dense environments at $z>3$ \citep{Kubo2021quiescent,McConachie_2022,Jin2024_cosmic_vine,Tanaka2024,Kakimoto2024,Ito2025,de_Graaff2025NatAs,Umehata2025cluster} and the red sequence has been revealed at $z>2$ \citep{Willis2020cluster,Ito2023,Tanaka2024}, enabling the study of potential environmental effects. More recently, top-heavy stellar mass functions (SMFs) have been found in observations of protoclusters at $z\gtrsim2.5$ \citep{Sun2024_top_heavy_cl1001,Sun2025_bigfoot,Galbiati2025} and in simulations \citep{Chartab2025_massive_qgs_smf}, indicating a top-down scenario of cluster galaxy formation.
With an increasing number of JWST images and spectra, we find ourselves at an ideal moment  to further push these frontiers and unveil the detailed processes that govern the formation and growth of galaxy clusters.

On the one hand, simulations predict that protoclusters consist of a series of dark matter halos of subgroups assembled along the cosmic web. 
Nevertheless, characterising dark matter halos with observations remains a challenging task because traditional tracers such as diffuse X-ray emission, the Sunyaev-Zel'dovich effect, and gravitational lensing are observationally expensive at high redshift \citep{Clowe2006,Wang_T2016cluster,Gobat2019,DiMascolo2023,Cha2025cluster,Zhou2025_SPT,Travascio2025}, leaving only stellar to halo mass relations (SHMR) as a practical tool for estimating dark matter masses for high-$z$ samples, although they come with a large uncertainty. Recently, the exploration of SHMR has been pushed towards higher redshift \citep{Shuntov2022,Paquereau2025} and stellar mass measurements have been significantly improved with deep multi-wavelength surveys \citep[e.g.][]{Weaver2022COSMOS2020}, allowing for better constraints to be set on the halo mass of $z\gtrsim3$ structures.
For example, \citet{Sillassen2024} extensively tested several methods of estimating halo masses in a sample of massive groups at $1.5<z<4$ using data from the Northern Extended Millimeter Array (NOEMA) and the COSMOS2020 catalogue \citep{Weaver2022COSMOS2020}, finding that the SHMR-based methods can constrain the halo mass uncertainty down to approximately $0.3\,{\rm dex}$ at $z\sim3$. 
With these improved methods and new JWST data, protoclusters in JWST deep fields offer ideal targets to characterise the assembly of massive dark matter halos.

In this paper, we use JWST data to investigate the galaxy population, SMFs, and dark matter halos in the $z=3.44$ protocluster Cosmic Vine found by \cite{Jin2024_cosmic_vine} in the JWST CEERS field.
In Sect. \ref{sec:data}, we present the observations, the sample selection, and the methodology. Section \ref{sec:res} describes the main results and discusses the implications. We summarise our results and present our conclusions in Sect. \ref{sec:conc}. 
We adopted a flat cosmology with $\Omega_m=0.27$, $\Omega_\Lambda=0.73$, and $H_0=70\,{\rm km\,s^{-1}\,Mpc^{-1}}$, and used an initial mass function (IMF) from \citet{Chabrier2003}. All reported magnitudes are in the AB system \citep{Oke1974_AB_sys}.

\section{Data and methodology}

\begin{figure*}[!htbp]
    \centering
    \includegraphics[width=0.69\textwidth]{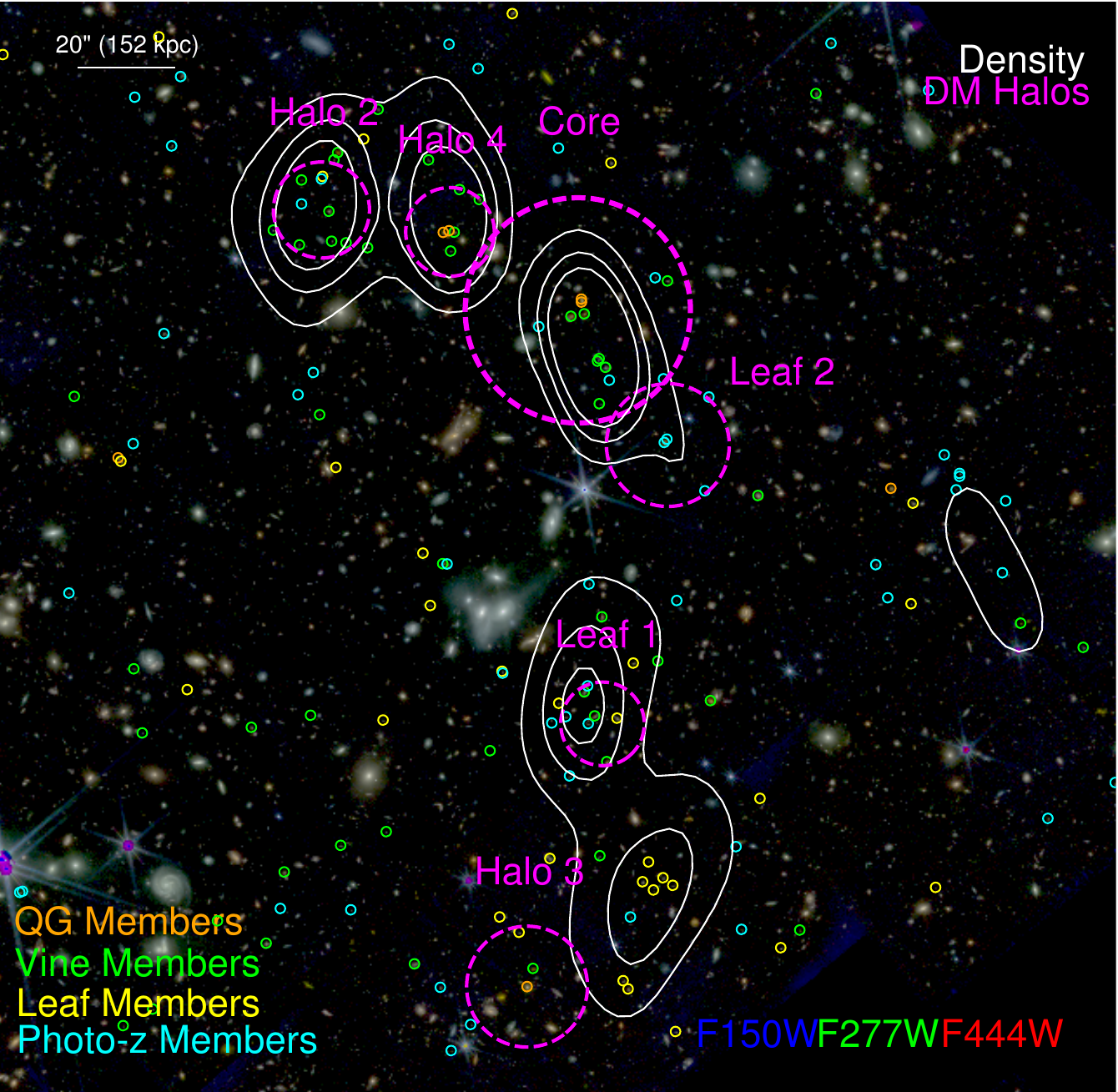}
    \includegraphics[width=0.2727\textwidth]{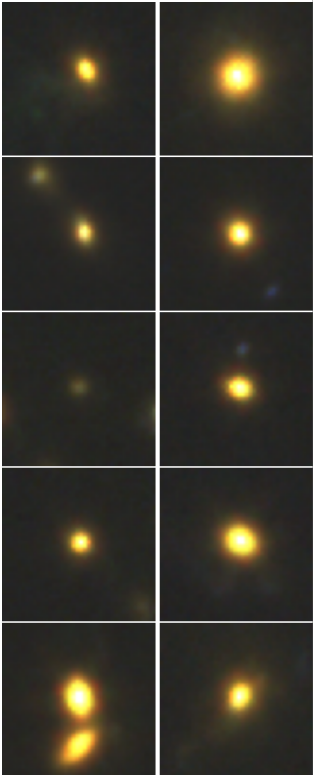}
    \caption{{\it Left:} Colour image of the densest region of the Vine and the Leaf. Colours correspond to JWST/F150W, F277W, and F444W as the blue, green, and red channels respectively. The projected density is shown as white contours at $2,3,~{\rm and}~4\,\sigma$ levels. The virial radii of the six most massive identified dark matter halos are marked with magenta dashed circles. Quiescent members of the Cosmic vine are marked with orange circles, spectroscopically confirmed members of the Cosmic Vine are marked with green circles, spec-z confirmed members of the Leaf are marked with yellow circles, and candidate members with cyan circles. {\it Right:} NIRCam/F150W, F277W, and F444W colour images of 11 quiescent galaxies in the Cosmic Vine. The field of view of each cutout is 2".}
    \label{fig:col-img}
\end{figure*}

\label{sec:data}
\subsection{DJA photometry and spec-z catalogues}
\label{sec:specz}
For the photometric data throughout the paper, we use JWST and Hubble Space Telescope (HST) data from the publicly available DAWN JWST Archive \citep[DJA; photometry v7.4; spec-z v4.4;][]{Valentino2023,brammer2025_DJA,DeGraaff2025,Heintz2025}. For the spectroscopic data, we made use of the DJA spectra compilation v4.4, using only robust (grade=3) spectra, in addition to the \cite{Jin2024_cosmic_vine} sample which includes spectra from Keck/MOSFIRE, MOSDEF, Keck/DEIMOS, and DEEP3 \citep{Schreiber2018_specz,Kriek2015,UrbanoStawinski2024_lya,Cooper2012}, which includes new NIRSpec data from the following JWST programs; AEGIS (PID \#1213, PI: Luetzgendorf), CEERS (PID \#1345, PI: Finkelstein; PID \#2565, PI: Glazebrook; PID \#2750, PI: Arrabal-Haro; PID \#4106, PI: Nelson; PID \#4287, PI: Mason), RUBIES (PID \#4233, PI: de Graaff),  CAPERS (PID \#6368, PI: Dickinson),  and DeepDive (PID \#3567, PI: Valentino; \citealt{Finkelstein2025,ArrabalHalo2023,DeGraaff2025,Ito2025deepdive}). Imaging data from the DJA used in this work includes data from CEERS \citep{Finkelstein2025} and CANDELS \citep{Grogin2011,Koekemoer2011}. We estimated the depth of the observations using several methods (Sect. \ref{sec:image_depths}), the imaging data includes the following passbands; HST/F435W; F606W; F814W; F105W; F125W; F140W; F160W and JWST/F115W; F150W; F182M; F200W; F210M; F277W; F356W; F410M; and F444W. A detailed description of DJA data reduction, calibration, and source extraction is presented in \citet{Valentino2023}. In short, we retrieved the level-2 products from the Mikulski Archive for Space Telescopes (MAST) and processed them with the \texttt{Grizli} pipeline \citep{Brammer2021,Brammer2022}. Images are aligned to Gaia DR3 \citep{Gaia2021_dr3}, co-added, and drizzed \citep{Fruchter2002} to a final pixel scale of 0.04". Sources are extracted using \texttt{SEP} \citep{Barbary2016} in circular apertures with a diameter of 0.7", corrected within an elliptical Kron aperture \citep{Kron1980}, with photometric errors measured in empty apertures. The extraction is performed on combined long wavelength JWST images (F277W+F356W+F444W). The extracted photometry was fit with \texttt{EAZY-PY} to estimate photometric redshifts in the interval $z=0-18$, using 13 templates from the Flexible Stellar Populations Synthesis code \citep[FSPS][]{Conroy2010_fsps}, described in \citet{Kokorev2022} and \citet{Gould2023}. We adopted the 50th percentile of the redshift probability density function (PDF(z)) as the photometric redshift. The spectroscopic data are a mix of low resolution ($R\sim100$) NIRSpec Prism and medium resolution ($R\sim1000$) NIRSpec GRISM, utilising the gratings; G235M and G395M, and were reduced with the \texttt{MsaExp} pipeline \citep{Brammer2023}. 1D spectra are fit using \texttt{MsaExp}, and the redshifts were determined using either Balmer absorption lines or emission lines including $\lambda\lambda[{\rm O{\tiny III}}]4959,5007$, H$\alpha$, H$\beta$, Pa$\varepsilon$, Pa$\delta$, and Pa$\gamma$. 

To determine the completeness of the DJA CEERS catalogue, following the method presented in \citet{Paquereau2025}, we used the F444W photometry and fit a power law to the observed magnitudes and determined where the observed magnitudes deviate from the power law by 20 percent, as shown in \cref{fig:completeness}.
We estimate the 80\% completeness of the F444W observations in the CEERS redshift selected field catalogue to be 28.8~AB mag, while the 80\% completeness in F444W is 27.1 mag for the Cosmic Vine members. It is assumed the catalogues are 100\% complete at magnitude brighter than 28.5~mag and 26.8~mag, respectively (\cref{fig:completeness}).

The accuracy of the photometric redshifts are determined by calculating the normalised median absolute deviation ($\sigma_{\rm NMAD}$, Eq. 3 in \citealt{Weaver2022COSMOS2020}).
The fraction of outliers, $\eta$, is defined as $\eta=|\Delta z|>0.15(1+z_{\rm spec})$, and the bias is defined as $b={\rm median}(\Delta z)$. We show the $z_{\rm phot}-z_{\rm spec}$ comparison in \cref{fig:photz_specz}, where we find $\sigma_{\rm NMAD}=0.017$, $\eta=5.32\%$, and $b=0.001$.

\subsection{Membership and completeness}
Using the CEERS catalogue from the DJA, we identify photometric candidate members by selecting galaxies with the 16th and 84th percentile $z_{\rm phot}$ within $z=3.44\pm0.25$, (i.e $z_{\rm phot,16th}>3.19 ~\& ~z_{\rm phot,84th}<3.69$). This selection was chosen to cover both peaks in the spec-$z$ distribution of CEERS (\cref{fig:zhist}), while accounting for photo-$z$ uncertainty (Sect. \ref{sec:specz}, Fig. \ref{fig:photz_specz}).

Spectroscopic members were selected as galaxies with $3.3<z_{\rm spec}<3.5$, encompassing the two identified peaks in the spec-$z$ distribution of CEERS (\cref{fig:zhist}). The redshift distribution of spectroscopically confirmed members of the Vine and the Leaf can be seen in \cref{fig:zhist}.

By enforcing the 16th and 84th percentile of the photometric redshift probability density function (PDF(z)) of each galaxy to be in a small threshold, we were able to limit the number of expected interlopers; however, we  also ended up introducing a bias toward brighter galaxies. This bias can be seen in \cref{fig:completeness}, where the 80\% completeness level of members is 1.7~mag brighter than the field. To estimate the mass completeness, we followed the method from \citet{Pozzetti2010}. In short, we selected the 30\% faintest galaxies above the completeness limit and rescaled their masses as if they were detected at the completeness limit following \cref{eq:massrescale}.
\begin{equation}
    \log M_{m_{\rm lim}}=\log M_\ast+0.4(m_{\rm F444W}-m_{\rm lim})
    \label{eq:massrescale}
.\end{equation}
Then, we defined the mass completeness limit as the 95th percentile of the $M_{m_{\rm lim}}$ distribution. This yields stellar mass completeness limits of $M_\ast>10^{8.32}\,{\rm M_\odot}$ for the field and $M_\ast>10^{8.61}\,{\rm M_\odot}$ for the Vine members.

We found 183 spec-$z$ confirmed galaxies and 314 photo-$z$ candidate members.
Based on these newly identified members, we updated the galaxy overdensity using the weighted adaptive kernel (WAK) mapping code from \citet{Brinch2023}. In short, WAK iteratively computes the galaxy surface density field taking photometric redshift uncertainty into account in fixed kernels at the positions of the galaxies, over a grid search of kernel widths from 0.0001-0.1 deg. In this work, the optimal global kernel width is $18.5"$ ($140.7\,{\rm pkpc}$). The updated overdensity contours are shown in Fig.~\ref{fig:col-img}.
We note that given the uncertainty of photometric redshifts (\cref{fig:photz_specz}), it is impossible to distinguish if photo-$z$ candidate members belong to the Vine or the Leaf, so we treated all candidate members as members of the whole structure.

\subsection{Bagpipes SED fitting}
Using a similar setup as described in \citet{Jin2024_cosmic_vine}, we fit the DJA JWST+HST photometry using Bagpipes \citep{Carnall2018_bagpipes}. For spectroscopically confirmed members, we fixed the redshift of the fit, while for candidate members, we let the redshift vary between $3<z_{\rm phot}<4$, finding overall consistent photometric redshifts as \texttt{EAZY}. Specifically, we used a double-power-law star formation history, allowing for stellar metallicities in the range $-2.3<\log(Z/{\rm Z_\odot})<0.7,$ with a log-uniform prior; an age of the Universe at star formation history turnover in the range $0.1\,{\rm Gyr}<\tau<15\,{\rm Gyr}$ and shape parameters with log-uniform priors in the ranges $-2<\log\alpha<4$ and $-2<\log\beta<4$; a dust attenuation law from \citet{Salim2018}, with attenuation in the range $0<A_V<4$; deviation from a \citet{Calzetti2000} curve in the range $-0.3<\delta<0.3,$ with a Gaussian prior $\mu_\delta=0$ and $\sigma_\delta=0.05$; a 2175~Å bump strength in the range $0<B<5$ and an extra attenuation of stars in birth clouds in the range $1<\eta<5$, along with nebular emission \citep[precomputed with CLOUDY][]{Ferland2017} with radiation fields in the range $-4<\log(U)<-2;$ and gas-phase metallicities in the range $-2.3<\log(Z_{\rm neb}/{\rm Z_\odot})<0.7$ with a log-uniform prior. All priors are uniform, unless otherwise specified.

\subsection{Quiescent galaxy selection}
\label{sec:qg_sel}
We classified quiescent galaxies as galaxies that either fulfil the rest-frame UVJ colour criteria from \citet[as detailed in our \cref{fig:uvj_red_sequence} (left)]{Williams2009} or that have an 84th percentile posterior specific star formation rate (${\rm sSFR}={\rm SFR}/M_\ast$) of $\log({\rm sSFR_{\rm 84th}}/{\rm yr^{-1}})<-10.3$ ($\sim1.8\,{\rm dex}$ below the star-forming main sequence at $z=3.44$, \citealt{Schreiber2015}). The adopted sSFR limit is conservative, compared to the criterion of \citet{Carnall2023}, (${\rm \log(sSFR/yr^{-1})}<0.2/t_{\rm H}$, where $t_{\rm H}$ is the age of the Universe at $z=3.44$), which would be $\log({\rm sSFR_{50th}}/{\rm yr^{-1}})<-10.0$ at the redshift of the Cosmic Vine. We present the results in Sect. \ref{sec:qg_res}. We note that while the completeness limit of quiescent galaxies is likely fainter than the SFG completeness limit \citep[e.g.][]{Weaver2022COSMOS2020}, the faintest QG member is 2 mag brighter than the SFG completeness limit in F444W. Furthermore, the least massive QG is 1 dex more massive than the SFG stellar mass completeness limit, where the difference in COSMOS2020 is at most 0.5 dex \citep{Weaver2022COSMOS2020}.

\subsection{Dark matter halo mass estimates}
\label{sec:darkmatter}
To estimate the dark matter halo masses, we followed the methods presented in \citet{Sillassen2024}. In short, we found halos using the 2D density contours (\cref{fig:col-img}) as initial guesses of halo centres. Given that the overdensity contours are only calculated with the 2D distribution of galaxies without weighting stellar mass, we thus complement the halo finding by adding the locations of massive galaxies in relatively low overdensity, such as Halo 3 and Leaf 2. We obtained an initial halo mass by scaling the most massive galaxy within a 15" circle (i.e. the virial radius of a $\log(M_{\rm h}/{\rm M_\odot})=12.7$ halo at $z=3.44$, the progenitor of a Virgo-like cluster, see \citealt{Chiang2013cluster}) with the stellar to halo mass relation (SHMR) from \citet[$M_{\rm h}(1)$\footnote{We use $M_{\rm h}(n)$ as short to refer to the n-th halo mass method presented in \citet{Sillassen2024}}]{Behroozi2013Mhalo}. Using the recovered initial halo mass, we infer a virial radius using the relation from \citet{Goerdt2010core}, and find a new halo centre by calculating the stellar mass weighted centre of mass of the galaxies within the new virial radius. With the updated centres, we calculated the total stellar mass, accounting for background contamination and stellar mass completeness \citep[e.g.][]{Daddi2021Lya,Daddi2022Lya}. We then scaled the total stellar mass with SHMRs from \citet[$M_{\rm h}(2a)$]{Shuntov2022} and \citet[$M_{\rm h}(2b)$]{van_der_Burg2014}. Other than using stellar masses, we estimated the overdensity of each halo and use this overdensity to estimate the halo mass ($M_{\rm h}(3)$). We did this iteratively, until the halo mass and halo members converge. The halo masses of the four methods are overall consistent; thus, we adopted an average of the four as the best estimate of halo mass.

Furthermore, we estimated halo mass of the core using the velocity dispersion of member galaxies in two ways, for both methods, we defined the core members as galaxies within the expected virial radius and $3\sigma_v$ of a $\log(M_{\rm h}/{\rm M_\odot})\sim13$ halo at $z=3.438$ for which we:\ (1)  measured the standard deviation of redshift by fitting the redshift distribution of the core galaxies with a Gaussian,  converted to a velocity dispersion with $\sigma_v=c\sigma_z/(1+z)$; and (2) following \citet{Bayliss2014}, we calculated the square root of the bi-weight variance of the galaxy proper velocities ($v_i=c(z_i-z)/(1+z)$) using jackknife sampling.


\section{Results and discussion}
\label{sec:res}

\subsection{Large-scale structure and subgroups}
\label{sec:lss_res}
With a total of 183 spectroscopic redshifts and 314 photo-$z$ members, we confirm the large-scale structure reported by \citet{Jin2024_cosmic_vine}. The new spec-$z$s significantly improve the membership completeness, enabling us to identify several sub-structures in the Vine. As shown in the spec-$z$ histogram in \cref{fig:zhist}, we find two major peaks at $z=3.44$ and $z=3.34$, of which the $z=3.44$ peak traces the main structure of the Vine. The $z=3.34$ peak reveals a foreground structure that is obviously more diffuse than the Vine, which we dubbed the Leaf. The new data confirm the massive and overdense nature of the Vine, while the Leaf is dominated by low-mass galaxies and consists of two substructures. The spec-$z$ confirmed members of the Vine span an area of $8.2\,{\rm pMpc}\times3.6\,{\rm pMpc}$.

As traced by the galaxy overdensities in \cref{fig:col-img}, we identified four subgroups in the Vine. In the core, four more galaxies were spectroscopically confirmed in addition to the sample from \citet{Jin2024_cosmic_vine}. 
The other subgroups also host massive galaxies ($\log(M_\ast/{\rm M_{\odot}})>10$), with two to ten spectroscopically confirmed members in each subgroup. 

In addition to spec-$z$ members, we selected 314 photo-$z$ members with $3.19<z_{\rm phot}<3.69$. Together, we have 497 confirmed and candidate  members that extend over the entire CEERS field, corresponding to $13.1\,{\rm pMpc}\times4.1\,{\rm pMpc}$. 
We note that the true size of the large-scale structure is likely extending further than the field of view (FoV) of the CEERS field, additional data are needed to verify if this is the case.

\subsection{Quiescent galaxies and red sequence}
\label{sec:qg_res}

\begin{figure*}[!htbp]
    \centering
    \includegraphics[width=\columnwidth]{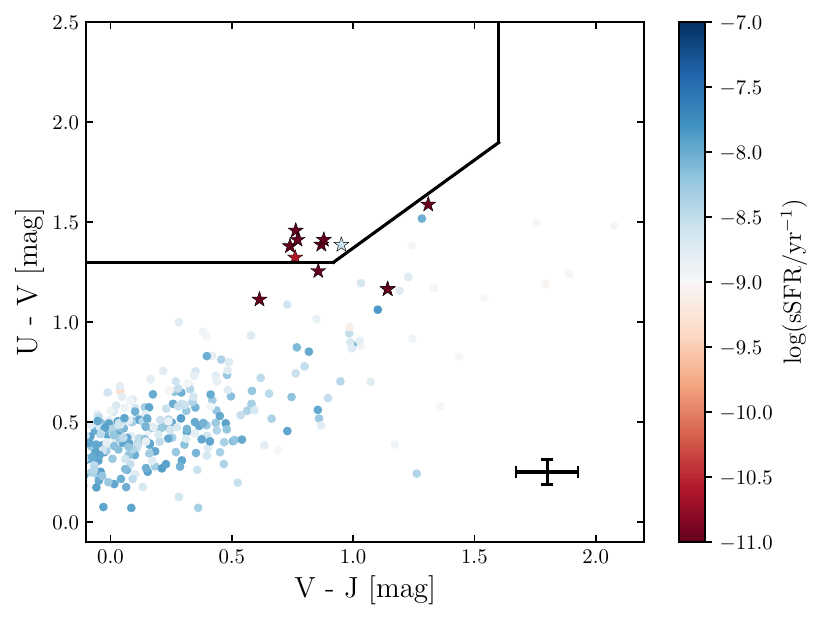}
    \includegraphics[width=\columnwidth]{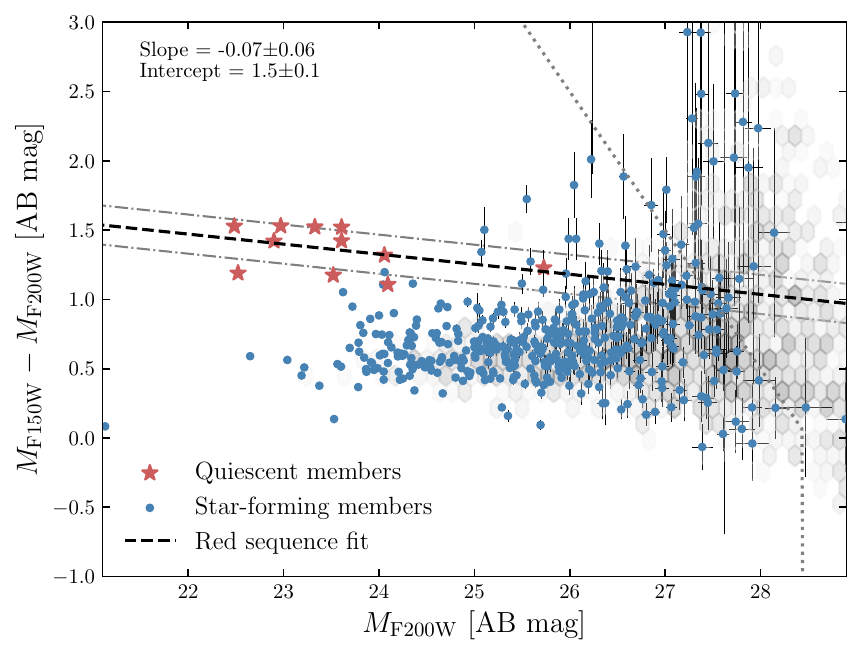}
    \caption{\textit{Left}: Rest-frame UVJ diagram of protocluster members. The colour selection function is shown in black lines. Identified quiescent galaxies are marked with a star. The sSFRs of the galaxies is coded in colour. The cross in the bottom right corner shows the average uncertainty \textit{Right}: Colour-magnitude diagram of quiescent (red stars) and star-forming (blue dots) members. The background hexagons show the density of field galaxies. Overlaid is a linear fit to the quiescent galaxies (black dashed) and the scatter (gray dot-dashed), with the slope and intercept at $M_{\rm F200W}=21$ shown in text. The observational depth is shown with a gray dotted line.}
    \label{fig:uvj_red_sequence}
\end{figure*}

Using the UVJ and sSFR criteria in Sect. \ref{sec:qg_sel}, we selected 11 quiescent galaxies (\cref{fig:col-img} right), in which 10 are within the Vine (5 spec-$z$ confirmed) and 1 spec-$z$ confirmed within the Leaf (the identified quiescent galaxies are shown in Fig.~\ref{fig:uvj_red_sequence} and Table~\ref{tab:qg_cat}); of these QGs, 7 were selected with both the UVJ and sSFR criteria and 4 were selected using only the sSFR criterion (\cref{fig:uvj_red_sequence,tab:qg_cat}). We note that all sources selected only with sSFR were confirmed to be quiescent with JWST spectra. As works in the literature  have demonstrated that the traditional UVJ selection might miss some quiescent galaxies (e.g. \citealt{Baker2025_expanded_uvj}), we also used the expanded UVJ selection criterion Eq. (3) in \citet{Baker2025_expanded_uvj} as a test. As a result, we found 10 of the 11 QGs are aptly selected. The only missing one can be selected with sSFR and has been spectroscopically confirmed as a QG by \cite{Jin2024_cosmic_vine}.
Two of the QGs are a merging pair already identified by \cite{Jin2024_cosmic_vine} and spectroscopically confirmed by \cite{Ito2025}, which are the central galaxies in the most massive core. The remaining sources were also identified as QGs by literature studies \citep{Merlin2019,Shahidi2020,Carnall2023_ceers_qgs,Valentino2023,Long2024}. Three QGs are located in the two dense subgroups just north-east of the core, while the six other identified QGs lie outside of the densest regions of the Vine.

We highlight that this is the largest sample of quiescent galaxies ever detected in a $z>3$ dense environment, as literature studies have only identified 1--7 QGs per structure at $z>3$ \citep{Kubo2021quiescent,McConachie_2022,Tanaka2024,Kakimoto2024,de_Graaff2025NatAs,Baker2025_qgs_overdensities}.
This large number of quiescent galaxies in a $z>3$ protocluster appears significantly higher than the field level. Thus, we constrained the number density of quiescent galaxies in the Vine, first  adopting a sky area of $8.2\times3.6$~Mpc$^2$ in physical scale (as reported in Sect. \ref{sec:lss_res}) and a redshift range of $3.4<z<3.5$, which corresponds to a co-moving volume of 4.86$\times10^4$~cMpc$^3$. We estimated the QG number density of the Vine with three different assumptions: (1) all nine QG candidates with $\log(M_*/{\rm M_\odot})>10$ are a part of the Vine; (2) inspired by the work of \citet{Long2024}, we performed $10^4$ simulations with MCMC sampling, assuming Gaussian photo-z uncertainties, adopting the inner 68th percentile of the posterior number density distribution and accounting for Poissonian noise; and (3) for the five photo-z selected QGs, we assumed their chance of being within the Vine (46\%) is consistent to the ratio between spec-z members within the Vine and all photo-z selected candidate members with a spec-z. 

We find (1) yields a number density of $1.85\times10^{-4}\,{\rm cMpc^{-3}}$ ($N_{QG}=9$); (2) yields $1.02^{+1.05}_{-0.23}\times10^{-4}\,{\rm cMpc^{-3}}$ ($\bar{N}_{QG}=5$); and (3) gives $1.3\times10^{-4}\,{\rm cMpc^{-3}}$ ($\bar{N}_{QG}=6.3$). Taking into account all three estimates, we obtain a number density in the range of $1.02-1.85\times\,10^{-4}\,{\rm cMpc^{-3}}$. 
This number density is $1.5-2.8$ times higher than the field density $6.57\pm1.03\times10^{-5}$~cMpc$^{-3}$ reported by \cite{Baker2025QGs} at $3.0<z<3.5$ and $\log(M_*/{\rm M_\odot})>10$, indicating an elevated abundance of quiescent galaxies in dense environment, which is further highlighted by the enhanced quiescent fraction compared with the field (see Sect. \ref{sec:smf_qf}). This elevated abundance of quiescent galaxies in the Cosmic Vine is likely the origin of the high abundance of quiescent galaxies at $3<z<4$ in the CEERS field reported by \citet{Carnall2023}, which was attributed to cosmic variance by \citet{Valentino2023}.

\begin{table*}[!htbp]
    \centering
    \renewcommand\arraystretch{1.2}
    \caption{Physical properties of quiescent members.}
    \begin{tabular}{c | c | c | c | c | c | c}
    \hline
        RA & DEC & $z$ & $\log({M_\ast/{\rm M_\odot}})$ & ${\rm\log (sSFR/{yr^{-1}})}$ & U-V & V-J\\
        $[{\rm deg}]$ & $[{\rm deg}]$ & -- & -- & -- & [mag] & [mag]   \\
        \hline
        214.87123 & 52.84507 & 3.437 & 10.8$\pm$0.1 & <-11.3 & 1.2$\pm$0.1 & 1.1$\pm$0.1$^{a,c,e,f}$\\
        214.86605 & 52.88426 & 3.433 & 10.8$\pm$0.1 & <-28.5 & 1.4$\pm$0.1 & 0.7$\pm$0.1$^{a,b,c,d,e,f}$\\
        214.86605 & 52.88409 & 3.443 & 10.8$\pm$0.1 & <-10.9 & 1.6$\pm$0.1 & 1.3$\pm$0.1$^{b,e,f,g}$\\
        214.98181 & 52.99124 & 3.43$\pm$0.05 & 10.8$\pm$0.1 & <-22.2 & 1.5$\pm$0.1 & 0.8$\pm$0.1$^{d,e,f,g}$\\
        215.06586 & 52.93295 & 3.55$\pm$0.10 & 10.7$\pm$0.0 & <-22.2 & 1.4$\pm$0.1 & 0.9$\pm$0.1$^{d,g}$\\
        214.83685 & 52.87346 & 3.23$\pm$0.07 & 10.6$\pm$0.1 & <-8.6 & 1.4$\pm$0.1 & 1.0$\pm$0.1$^{f,g}$\\
        214.76725 & 52.81770 & 3.65$\pm$0.10 & 10.5$\pm$0.1 & <-14.1 & 1.4$\pm$0.1 & 0.9$\pm$0.1$^{d,e,f,g}$\\
        214.95789 & 52.98031 & 3.54$\pm$0.10 & 10.4$\pm$0.1 & <-22.3 & 1.4$\pm$0.1 & 0.8$\pm$0.1$^{e,g}$\\
        214.87910 & 52.88806 & 3.442 & 10.3$\pm$0.1 & <-10.5 & 1.3$\pm$0.1 & 0.8$\pm$0.1$^{c,d,f,g}$\\
        214.87856 & 52.88816 & 3.450 & 9.6$\pm$0.1 & <-11.8 & 1.1$\pm$0.1 & 0.6$\pm$0.1$^{f}$\\
        214.90955 & 52.87503 & 3.353 & 10.0$\pm$0.1 & <-12.1 & 1.3$\pm$0.1 & 0.9$\pm$0.2$^g$$\dagger$\\
        \hline
    \end{tabular}
    { \\Source identified as QG in $^a$\citet{Jin2024_cosmic_vine}, $^b$\citet{Ito2025}, $^c$\citet{Merlin2019}, $^d$\citet{Shahidi2020}, $^e$\citet{Carnall2023_ceers_qgs}, $^f$\citet{Valentino2023}, $^g$\citet{Long2024}. $\dagger$ Member of the Leaf.}
    \label{tab:qg_cat}
\end{table*}

With a sample of 11 quiescent galaxies, we investigated whether the ubiquitous red sequence of local galaxy clusters is already present in this early large-scale structure. To do so, we created a colour-magnitude diagram using NIRCam F150W/F200W colour, where the two bands respectively cover blue- and red-wards of the Balmer and $D_n{\rm 4000}$ break at $z=3.44$. In \cref{fig:uvj_red_sequence} (right), we fit the colour-magnitude of quiescent members with a linear relation normalising at $M_{\rm F200W}=21$, resulting in a negative slope $-0.07\pm0.06$ and an intercept of $1.5\pm0.1$. This red sequence is in agreement with the observed red sequence intercept ($1.41$) and slope ($-0.013$) of the $z=1.98$ cluster XLSSC122 \citep{Willis2020cluster}.

This red sequence is one of the two earliest known to date: the other one is at $z\sim4$ reported by \citet{Tanaka2024}, which exhibits a red sequence in $H-K$ vs $K$ diagram but is more scattered than the Vine's. We note that the structure reported by \citet{Tanaka2024} is more diffuse and less complete in terms of spectroscopically confirmed members than the Vine.

The early emergence of a red sequence and the high abundance of quiescent galaxies suggest that the Cosmic Vine is in an early stage of forming typical cluster galaxies such as those ones in the Local Universe. 
As is evident from the multiple subgroups, it demonstrates that the red sequence arises before the subgroups coalesce and the structure virialises.
Intriguingly, only two massive quiescent members are in the core and the majority of them are located outside of the densest regions. These massive quiescent galaxies outside the dense core might have experienced pre-processing in subgroups, such as the Cosmic Rose \citep{Alberts2024_cosmic_rose}, which have since coalesced into single massive galaxies \citep[e.g.][]{Jin2023_cggz5}. This pre-processing could be a substantial environmental effect on the formation and overabundance of massive galaxies in the Cosmic Vine. However, as high pressure hot gas is unlikely to be present in such a large elongated structure, it disfavours the dominance of environmental quenching mechanisms \citep{Alberts2022} that require hot gas and virialisation, such as ram pressure stripping (RPS). Instead, these massive galaxies are likely dominated by mass quenching mechanisms, such as AGNs and compact starbursts \citep{Jin2024_cosmic_vine,Wang2024Natur}, indicating that they share similar quenching pathways with massive field quiescent galaxies \citep{Carnall2023,Ito2025,Valentino2025outflow}.

\subsection{Stellar mass function and quiescent fraction}
\label{sec:smf_qf}
To further characterise the assembly of the galaxy population in the Vine, we establish the SMFs of star-forming galaxies (SFGs) and QGs separately and fit them with the Schechter function (\citealt{Schechter1976LF}, \cref{eq:Schecter}). We used a Markov chain Monte Carlo (MCMC) fitting procedure for both members and the rest of the CEERS field. We selected a sample of 2863 field galaxies in CEERS as a control sample, defined as galaxies with $3<z_{\rm phot}<4$ that have not been selected as candidate members of the Vine.
The SMFs are constructed as

\begin{align}
\begin{split}
    \Phi d\log M=&\ln(10)\exp\left(-10^{\log M-\log M^\ast}\right)\\ 
    &\times\left(\Phi^\ast10^{\log M-\log M^\ast}\right)^{\alpha+1}d(\log M). \label{eq:Schecter}
\end{split}
\end{align}

We calculated the observed $\Phi d(\log M)$, following the $1/V_{\rm max}$ method described in \citet{Schmidt1968_smf} and \citet{Weigel2016}. In short, for each galaxy in each mass-bin we estimate the maximum volume in which a galaxy at the given mass and redshift could be detected; $V_{\rm max,i}$. The $V_{\rm max,i}$ for each galaxy was calculated with the expression\begin{equation}
    V_{\rm max,i}=\frac{4\pi}{3}\frac{\Omega_{\rm survey}}{\Omega_{\rm sky}}\left(d_c(z_{\rm max})^3-d_c(z_{\rm min})^3\right)
    \label{eq:vmax}
,\end{equation}
where $\Omega_{\rm survey}$ is the sky coverage of the survey ($105.2\,{\rm arcmin^2}$), and $\Omega_{\rm sky}=41253\,{\rm deg^2}$ is the total area of the sky, $d_c(z)$ is the co-moving distance to redshift $z$, $z_{\rm min}$ is the lower edge of the redshift bin, and $z_{\rm max,i}$ is the minimum value of the upper edge of the redshift bin and the maximum redshift at which a galaxy at this mass could be detected; $z_{\rm max,i}=\min(z_{\rm max,bin},z_{\rm max,galaxy})$. The number density in each mass bin is then weighted by the JWST/F444W completeness of each galaxy and ${V_{\rm max,i}}$,
\begin{equation}
    \Phi d\log M=\sum_{n=1}^{N_{bin}}\frac{w_{\rm complete,i}}{V_{\rm \max,i}}
.\end{equation}
The SMF of the Vine is normalised to the field SMF, based on the number of galaxies in each sample.

\begin{figure*}[!htbp]
    \centering
    \includegraphics[width=\columnwidth]{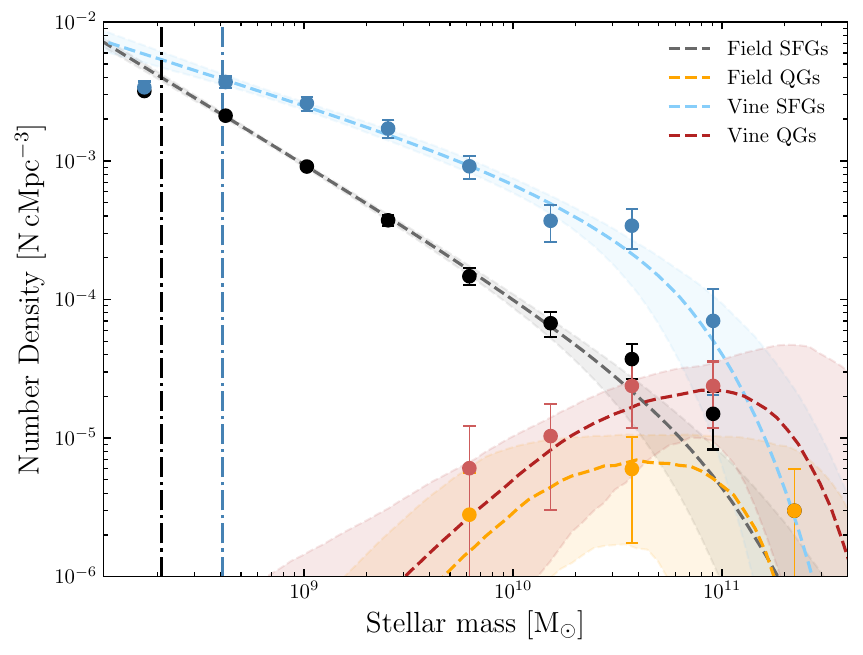}
    \includegraphics[width=\columnwidth]{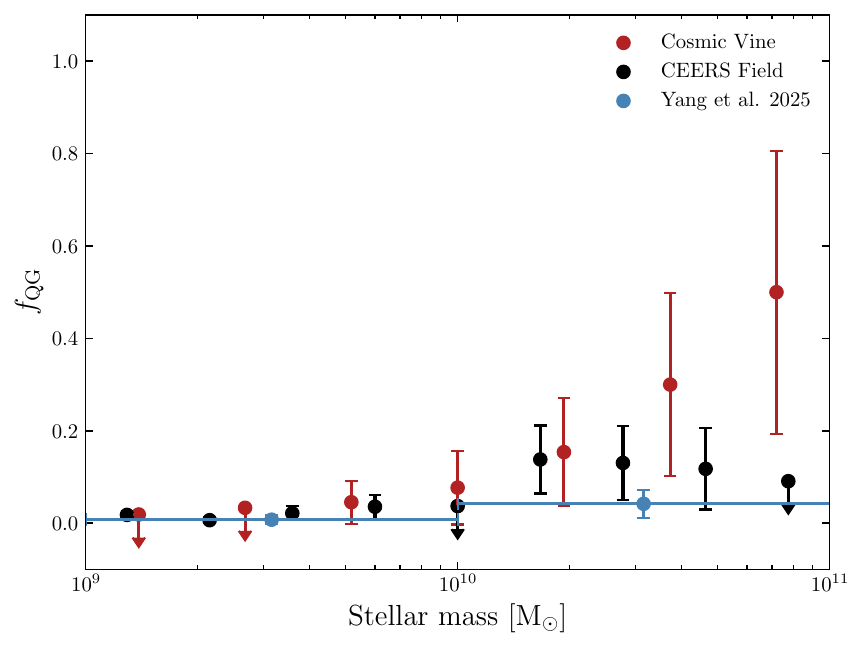}
    \caption{Stellar mass functions (SMFs) and quiescent fractions. 
    {\it Left:} SMFs for the star-forming (light blue) and quiescent (red) members of the Vine, fitted with a Schechter function. The black dots and line show the CEERS SFG field sample, and the orange dots and line show the CEERS QG sample. The vertical lines mark the stellar mass completeness limit of the field (black) and Cosmic Vine members (blue). 
    {\it Right:} Quiescent fractions of Cosmic Vine members (red) and the rest of the CEERS field (black). Blue dots with error bars represent the quiescent fractions from the JWST/PRIMER fields at $z\sim3.45$ \citep{Yang2025}.}
    \label{fig:smf_qg_fractions}
\end{figure*}

The measured and Schechter-fitted SMFs of the Cosmic Vine and CEERS field are shown in \cref{fig:smf_qg_fractions}-left, and the shape parameters of the fitted SMFs are shown in \cref{tab:smf_results}, we note that only data points above the stellar mass completeness limit are considered in the fitting procedure. If the galaxies belonging to the Leaf are excluded from the SMF fit, the parameters do not change significantly ($\log(M^*_{\rm no\,leaf}/{\rm M_\odot})=10.9^{+0.6}_{-0.4},\alpha_{\rm 1,no\,leaf}=-1.53^{+0.08}_{-0.07}$). The fitted $M^*$ is not significantly different between the CEERS field and Cosmic Vine members; however, $\alpha$ is significantly higher in the Cosmic Vine compared with the field, indicating an enhancement of high-mass ($M_*>10^{10}\,{\rm M_\odot}$) star-forming galaxies. If we do not remove members of the Vine and fit the field SMF, we find $\log(M_{\rm \ast,Field+Vine}/{\rm M_\odot})=11.3^{+0.3}_{-0.3}$ and $\alpha_{\rm Field+Vine}=-1.83^{+0.03}_{-0.03}$, leading to the same conclusion.
Comparing to the literature results, the shape of the Vine SMF is similar to the top-heavy SMFs of the $z=2.51$ CL J1001 cluster in the COSMOS field \citep{Sun2024_top_heavy_cl1001} and the $z=3.98$ Bigfoot \citep{Sun2025_bigfoot}, which both show a high abundance of SFGs with $M_*=10^{10.5-11}{\rm M_\odot}$.

\begin{table}[!htbp]
    \centering
    \caption{Best-fit SMF parameters.}
    \renewcommand\arraystretch{1.4}
    \begin{tabular}{c | c | c}
         & $\log{(M^\ast/{\rm M_\odot})}$ & $\alpha$\\
        \hline
       Field  & $11.0^{+0.4}_{-0.4}$ & $-1.91^{+0.05}_{-0.04}$\\
       Vine SFGs single & $10.8^{+0.5}_{-0.4}$ & $-1.49^{+0.09}_{-0.07}$\\
       Vine QGs & $10.7^{+0.5}_{-0.4}$ & $0.5^{+1.1}_{-0.7}$\\
    \end{tabular}
    \label{tab:smf_results}
\end{table}

To investigate the environmental dependence of quenching galaxies, we measured the quiescent fraction $f_{\rm QG}=N_{\rm QG}/(N_{\rm QG}+N_{\rm SFG})$ in both the Vine and the rest of the CEERS field.
As shown in \cref{fig:smf_qg_fractions} (right), the Vine exhibits an increasing $f_{\rm QG}$ at high stellar mass $\log(M_*/{\rm M_\odot})>10.5,$ compared with the rest of the field. 

Notably, at the most massive stellar bin $\log(M_*/{\rm M_\odot})\sim10.8$, the quiescent fraction of the field level is very low $f_{\rm QG}<0.09$, while in the Vine this mass bin is dominated by the two central merging quiescent galaxies \citep{Ito2025} with $f_{\rm QG}=50\pm30\%$. This again demonstrates the massive members of the Cosmic Vine are more evolved than field counterparts. An enhanced quiescent fraction was reported by \cite{McConachie_2022} in the $z=3.37$ MAGAZ3NE J0959 protocluster, which shows similar results as in this work.

\subsection{Stellar mass-size relation}
\begin{figure}[!h]
    \centering
    \includegraphics[width=\linewidth]{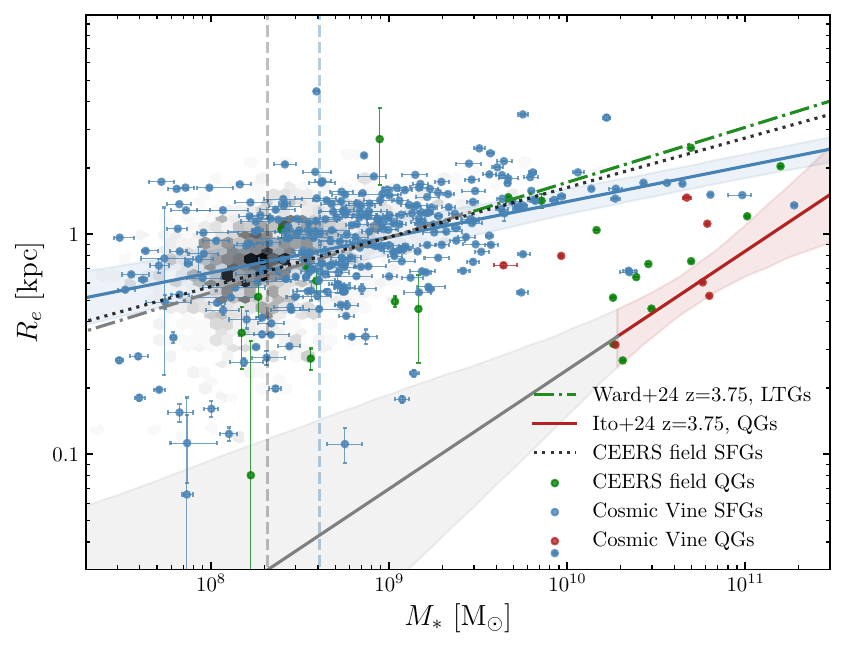}
    \caption{Mass-size relation of star-forming members (blue), quiescent members (red), field SFGs (black), and field QGs (green). Power-law fits to SFG members and field SFGs are shown with blue solid and black dotted lines respectively, with the fit uncertainty of SFG members shown as shaded region. Mass-size relations of late type galaxies from \citet[green line]{Ward2024_Mass_size_relation} and quiescent galaxies from \citet[red line]{Ito2024_QG_mass_size} are overplotted, the relations are extrapolated for SFGs with $M_\ast<10^{9.5}\,{\rm M_\odot}$ and QGs with $M_\ast<10^{10.3}\,{\rm M_\odot}$. The vertical dashed lines show the mass completeness limit of the CEERS field (gray) and Cosmic Vine (blue).}
    \label{fig:mass_size}
\end{figure}

To investigate whether there is a difference in the sizes of Cosmic Vine members and the rest of the field, we established the mass-size relation of star-forming and quiescent members using the morphology catalogue from the DJA \citep{Genin2025_DJA_morph}, where the size of each galaxy is fitted assuming identical radii across all JWST bands. We fit the mass-size relation to the functional form, with the results shown in Fig.~\ref{fig:mass_size}, following
\begin{equation}
    R_e/\,{\rm kpc}=A\left(\frac{M_\ast}{5\times10^{10}\,{\rm M_\odot}}\right)^{\alpha}
    \label{eq:mass-size}
,\end{equation}
where $A$ is the effective radius at $M_\ast=5\times10^{10}\,{\rm M_\odot}$ in kpc.

\begin{figure*}[!h]
    \centering
    \includegraphics[height=0.27\textheight]{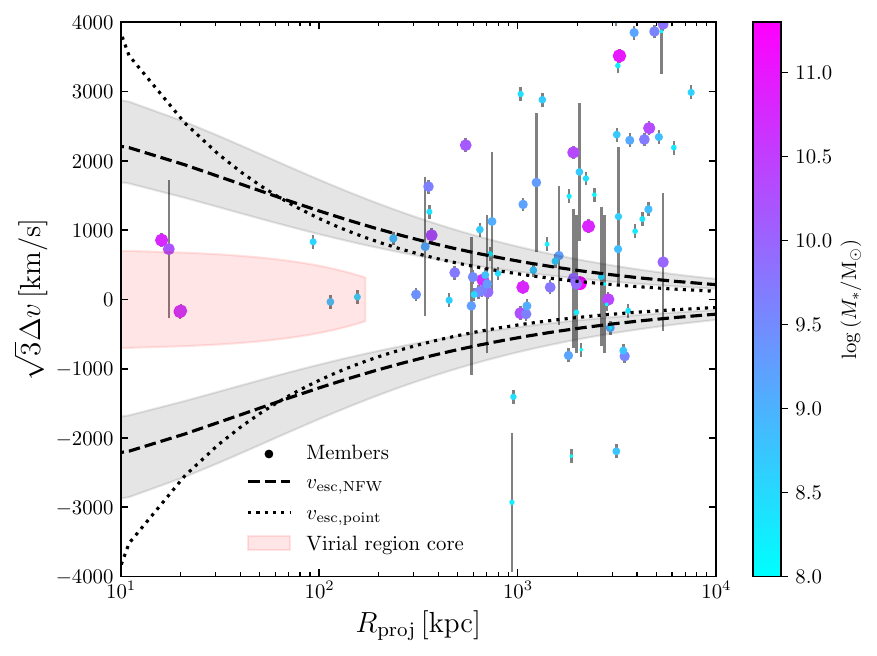}
    \includegraphics[height=0.26\textheight]{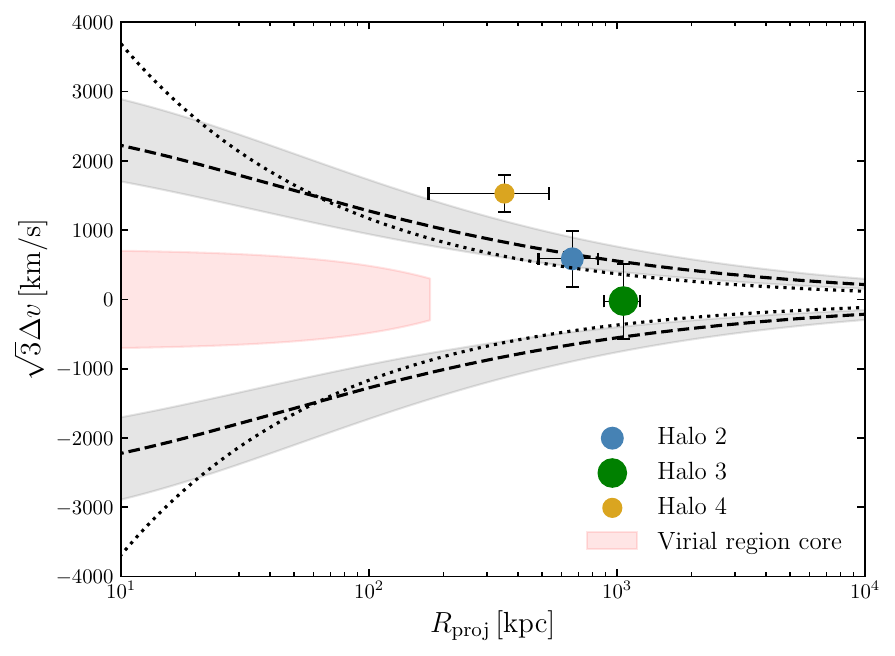}
    \caption{Phase-space diagrams for spectroscopically confirmed members and subgroups. {\it Left:} Galaxies are sized and coloured according to the stellar mass. Measured velocity differences are multiplied with a factor of $\sqrt{3}$, to account for measuring projected velocities \citep{Rhee2017}. 
    {\it Right:} Phase-space diagram for the subgroups with sizes coded with the estimated halo mass (\cref{tab:halos}). The errorbars on the subgroups show the estimated virial radius in the radial direction, and the uncertainty on the central velocity in the velocity direction. Overlaid: $v_{\rm esc,NFW}$ calculated as a NFW profile with $c_{\rm halo}=20.46$ shown as a black dashed line, and $v_{\rm esc,point}$ as a point mass distribution ($c_{\rm halo}\rightarrow\infty$) shown as a black dotted line. The virial region of the core is defined as $v_{\rm r} < v_{\rm r,crit} ~\&~ R<R_{\rm vir}$ from \citet*{Sanchis2004}, which is shown with a red shaded area.}
    \label{fig:phase_space}
\end{figure*}

As shown in Fig.~\ref{fig:mass_size}, SFGs in the CEERS field follow well the mass-size relation of $z\sim3.75$ late-type galaxies from \citet{Ward2024_Mass_size_relation}, while Cosmic Vine members display a slightly shallower slope. This results in a radius of 0.09 ($\sim2\sigma$) to 0.15 ($\sim3\sigma$) dex lower in star-forming Vine members with mass $\log(M_\ast/{\rm M_\odot})=10.7$ when compared to the rest of the field and the relation from \citet{Ward2024_Mass_size_relation} respectively (\cref{fig:mass_size}). This indicates that massive star-forming Vine members are in general more compact than field SFGs at the same redshift. The majority of the massive SFGs show a bulge-like core with a star-forming disk and high bulge-to-total ratios $B/T>0.5$, indicating that their compact sizes are likely due to rapid growth of bulges (e.g. \citealt{Huertas-Company2025,Shuntov2025morphology}).
Overall, QG members  follow the mass-size relation of $z\sim3.75$ QGs from \citet{Ito2024_QG_mass_size} with large scatter. However, the sample size of QGs remains too small to draw any conclusion on the size difference of QG members. 

\subsection{Dark matter halos and phase-space analysis}

Using the methods described in Sect. \ref{sec:darkmatter}, we estimated the dark matter halo mass of the six subgroups, with the best halo mass estimates presented in \cref{tab:halos}. 
As an alternative approach, we also measured the velocity dispersion of spec-z members in the core halo and derived its halo mass using Eq. (1) of \citet{Wang_T2016cluster}. We obtained (1): $\sigma_v=432\pm174\,{\rm km/s}$ and (2): $\sigma_v=427\pm184$, corresponding to halo masses of $\log(M_{\rm h}/{\rm M_\odot})=13.3\pm0.5$ and $\log(M_{\rm h}/{\rm M_\odot})=13.3\pm0.6$. This halo mass is in agreement with the average of the four other methods (\cref{tab:halos}). The large uncertainty on the velocity dispersion, and thereby halo mass, arises from the low statistic of spec-z confirmed galaxies in the core. 

\begin{table}[!htbp]
    \centering
    \setlength{\tabcolsep}{4pt}
    \renewcommand{\arraystretch}{1.2}
    \caption{Centres, masses, and central redshift of identified halos. }
    \begin{tabular}{c | c | c | c | c}
    \hline
        Name & RA & Dec & $\log(M_{\rm h}/{\rm M_\odot})$ & $z_{\rm center}$\\
        -- & $[{\rm deg}]$ & $[{\rm deg}]$ & -- & --\\
        \hline
        Core & 214.866 & 52.884 & $13.2\pm0.3$ & $3.436\pm0.006$\\
        Halo 2 & 214.890 & 52.889 & $12.1\pm0.3$ & $3.441\pm0.006$\\ 
        Halo 3 & 214.871 & 52.845 & $12.4\pm0.4$ & $3.436\pm0.008$\\
        Halo 4 & 214.878 & 52.888 & $12.0\pm0.5$ & $3.449\pm0.004$\\
        Leaf 1 & 214.866 & 52.861 & $11.9\pm0.6$ & $3.37\pm0.04$\\
        Leaf 2 & 214.858 & 52.876 & $12.4\pm0.5$ & $3.346\pm0.005$\\
    \hline
    \end{tabular}
    \label{tab:halos}
    \tablefoot{The expected virial radii of the halos are visualised in \cref{fig:col-img}.}
\end{table}

The large number of spec-z confirmed members across the entire structure allow us to create a phase-space diagram and assess whether the members and subgroups are gravitationally bound. To construct the phase-space diagram, we first measured the projected radii from the centre of the most massive core to the spec-z confirmed galaxies in physical kpc. Then we measured the velocity difference between to the redshift of central galaxies (\cref{tab:halos}), and the galaxies with $\Delta v_i=c(z_i-z_{\rm halo})/(1+z_{\rm halo})$. The velocity differences are multiplied by $\sqrt{3}$ to correct the measured projected velocities to 3D velocities \citep{Rhee2017} assuming velocity isotropy. Finally, we calculate the escape velocity as a function of distance from the central halo, using a Navarro-Frenk-White \citep*{Navarro1996} density profile as described in \citet{Rhee2017}, shown in \cref{eq:nfw_profile}, where $c_{\rm halo}$ is the concentration parameter of the dark matter halo. We estimated the $c_{\rm halo}$ using the relation from \citet{Ludlow2016}, expressed as

\begin{equation}
    \label{eq:nfw_profile}
    v_{\rm esc}=\sqrt{\frac{2GM_{\rm vir}}{R_{\rm vir}}\frac{\frac{\ln\left(1+c_{\rm halo}\frac{R}{R_{\rm vir}}\right)}{\frac{R}{R_{\rm vir}}}}{\ln(1+c_{\rm halo})-\frac{c_{\rm halo}}{1+c_{\rm halo}}}}
.\end{equation}

As shown in the phase-space diagram in \cref{fig:phase_space} (left), the members of the core are gravitationally bound to the most massive dark matter halo. Interestingly, we also found several massive galaxies with the same redshift as the core out to a projected radius of $R_{\rm proj}\sim3\,{\rm pMpc}$. This projected radius corresponds to a radius of $\sim15\,{\rm cMpc}$, aligning with the expected effective radius of a Coma-like cluster progenitor at $z\sim3.5$ \citep{Chiang2013cluster}. The most massive halo mass estimated at $\log(M_{\rm h}/{\rm M_\odot})=13.2\pm0.3$ is also consistent with the expectation of a Coma-like progenitor at $z\sim3.5$. Together with the size and mass of the Cosmic Vine, all evidence suggests it will evolve to a massive galaxy cluster with $\log(M_{\rm h}/{\rm M_{\odot}})>15$ at redshift $z\sim0$. 
Furthermore, we show the centres of each of the identified subgroups in \cref{fig:phase_space} (right). Two subgroups are enveloped within the escape velocity curves, while the third subgroup (Halo 4) is just above the velocity curve. This could indicate that the three subgroups are infalling clumps, which will eventually merge with the core. We note that massive galaxies not bound to the core are a part of the filamentary large structure that connects the core and subgroups of the Cosmic Vine.

We also note that the 3D separation between the core of the Cosmic Vine and the Leaf is $\sim87.4\,{\rm cMpc}$. While this distance is larger than the expected extent of a single forming galaxy cluster at $z\sim3.5$ \citep{Muldrew2015cluster,Chiang2017cluster}, it is well within the expected size of a filamentary structure of galaxy protoclusters similar to the Hyperion super-protocluster \citep{Cucciati2018,Ata2022cluster}.


\section{Conclusions}
\label{sec:conc}

The Cosmic Vine is one of the richest structures discovered at $z>3$ to date. In this work, we further revealed the large structure and studied the member galaxy populations and the dark matter halos. We have drawn the following conclusions:
\begin{enumerate}
    \item Using JWST archival data, we spectroscopically confirmed 136 members in the Cosmic Vine at $z\approx3.44$, and 47 members in a newly discovered foreground structure at $z\approx3.34,$ which we dubbed the Leaf. In total, we identified 314 candidate, and 183 confirmed members belonging to the whole structure, covering an area of $8.2\times3.6\,{\rm Mpc^2}$. 
    We revealed six subgroups with halo mass $\log(M_{\rm h}/{\rm M_\odot})\gtrsim12$, in which four of them are within the Vine and the remaining two subgroups belongs to the Leaf.

    \item Using both UVJ and sSFR, we identified 11 quiescent members with $\log(M_\ast/{\rm M_\odot})=9.5-11.0$ in the whole structure, constituting the largest sample of quiescent galaxies in overdense environments at $z>3$ to date. This large number of quiescent galaxies in a dense structure infers a high number density of quiescent galaxies $\sim1-2\times10^{-4}$~cMpc$^{-3}$ that is two to three times above the field level at $\log(M_*/{\rm M_\odot})>10$ \citep{Baker2025QGs}, demonstrating accelerated galaxy evolution in dense environment.
    
    \item Remarkably, the quiescent galaxies form a tight red sequence on the colour-magnitude diagram, significantly elevated from the star-forming members and the field galaxies. It is distinguished as one of the earliest red sequences known to date.

    \item We find robust top-heavy SMFs for both star-forming and quiescent members of the Vine compared to the field counterparts, while the quiescent fraction is significantly elevated at high stellar mass $\log(M_\ast/{\rm M_\odot})>10.5$. 
    
    \item Our stellar mass-size analysis shows that star-forming members are more compact than the field counterparts at higher stellar masses, which is consistent with the predictions from simulations \citep[e.g,][]{Sommer-Larsen2010_expected_compact_massive_galaxies}.
    
    \item The halo mass of the most massive subgroup, identified as the core of the structure in \citet{Jin2024_cosmic_vine}, is estimated to be $\log(M_{\rm h}/{\rm M_\odot})=13.2\pm0.3$. For the other five subgroups, we estimated halo masses in the range $\log(M_{\rm h}/{\rm M_\odot})=11.9-12.4$.
    
    \item The phase-space analysis of the structure reveals that several massive galaxies are gravitationally bound to the core, while members out to a distance of $R\sim3\,{\rm pMpc}$ have the same redshift as the core, consistent with the effective radius of coma-like cluster progenitors in simulations. Furthermore, the three identified subgroups of the Vine are possibly infalling clumps, supporting the scenario that these halos are merging and on the way to forming a massive cluster.
\end{enumerate}

 Coupling the evolved galactic population with the estimated mass and extent of the structure, the Cosmic Vine is likely evolving to a massive galaxy cluster with $\log(M_{\rm h}/{\rm M_\odot})>15$ at $z\sim0$. It already appears to be in an early stage of maturing, prior to coalescing into a single dark matter halo. With the wealth of data already available in CEERS and more incoming, the Cosmic Vine will be an ideal laboratory to study different aspects of cluster formation and galaxy evolution in the future.

\begin{acknowledgements}
We thank the anonymous referee for their constructive feedback. The Cosmic Dawn Center (DAWN) is funded by the Danish National Research Foundation under grant DNRF140. 
SJ acknowledges financial support from the European Union’s Horizon Europe research and innovation program under the Marie Skłodowska-Curie grant No. 101060888. GEM and SJ
acknowledge the Villum Fonden research grants 37440 and 13160. FV and KI acknowledge support from the Independent Research Fund Denmark (DFF) under grant 3120-00043B. This work is based on observations made with the NASA/ESA/CSA \textit{James Webb} Space Telescope. The data were obtained from the Mikulski Archive for Space Telescopes at the Space Telescope Science Institute, which is operated by the Association of Universities for Research in Astronomy, Inc., under NASA contract NAS 5-03127 for JWST. The specific observations analyzed can be accessed via \href{https://doi.org/10.17909/0bq4-kj71}{10.17909/0bq4-kj71}. These observations are associated with programs GTO \#1213; ERS \#1345; GO \#2565; DD \#2750; GO \#3567; GO \#4106; GO \#4233; GO \#4287; and GO \#6368. The authors acknowledge the teams and PIs for developing their observing program with a zero-exclusive-access period.
\end{acknowledgements}

\bibliographystyle{bibtex/aa}
\bibliography{biblio}

\begin{appendix}
\FloatBarrier
\section{Depths of observations}
\label{sec:image_depths}
\begin{figure}[!htbp]
    \centering
    \includegraphics[width=\linewidth]{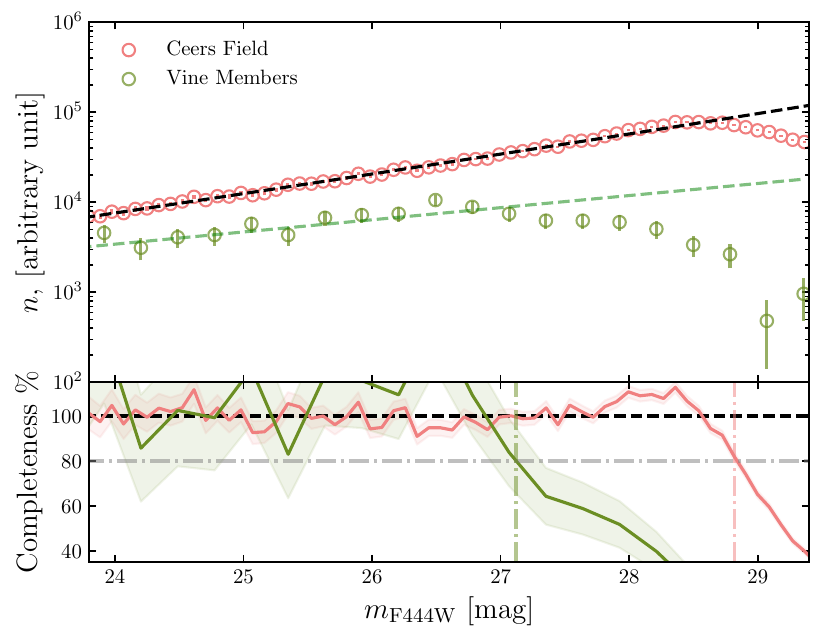}
    \caption{JWST/F444W Completeness of the entire CEERS field (red), and the confirmed and candidate members of the Cosmic Vine (green).}
    \label{fig:completeness}
\end{figure}

We estimated the $5\sigma$ depths of the observations using four methods:
(1) using the depth where the observed magnitudes deviate from a power-law by 20\% (Sect. \ref{sec:specz}, \ref{fig:completeness}); (2) measuring the magnitude of the faintest $5\sigma$ detected galaxy in each band in the catalogue; (3) fitting the negative flux distribution of all pixels in the images with a gaussian, and accounting for the number of pixels covered by a single aperture; and (4) placing random apertures in the images, avoiding bright sources, and fitting the flux distribution with a gaussian.

\begin{table}[!htbp]
    \centering
    \caption{$5\sigma$ depths measured in $0.7"$ apertures. }
    \renewcommand\arraystretch{1.25}
    \begin{tabular}{c|c|c|c|c}
        Filter & (1) & (2) & (3) & (4) \\
        \hline
        F435W & 28.5 & 28.2 & 27.9 & 27.6\\
        F606W & 28.7 & 28.1 & 28.0 & 28.0\\
        F814W & 28.5 & 27.8 & 27.8 & 27.8\\
        F105W & 27.7 & 27.5 & 27.9 & 27.0\\
        F125W & 27.9 & 27.2 & 27.9 & 27.2\\
        F140W & 27.5 & 26.4 & 27.3 & 26.6\\
        F160W & 27.9 & 27.2 & 28.1 & 27.3\\
        F115W & 28.6 & 28.9 & 28.5 & 27.8\\
        F150W & 28.3 & 28.8 & 28.3 & 27.6\\
        F182M & 28.3 & 28.5 & 28.0 & 27.5\\
        F200W & 28.3 & 29.1 & 28.5 & 27.9\\
        F210M & 28.1 & 28.2 & 27.9 & 27.2\\
        F277W & 28.8 & 29.3 & 29.1 & 28.5\\
        F356W & 29.0 & 29.4 & 29.1 & 28.6\\
        F410M & 28.5 & 28.5 & 28.3 & 27.9\\
        F444W & 29.0 & 28.9 & 28.7 & 28.4\\
    \end{tabular}
    
    \label{tab:placeholder}
\end{table}

\FloatBarrier

\section{Completeness}

\section{Spectroscopic redshifts}
\begin{figure}[!htpb]
    \centering
    \includegraphics[width=0.9\columnwidth]{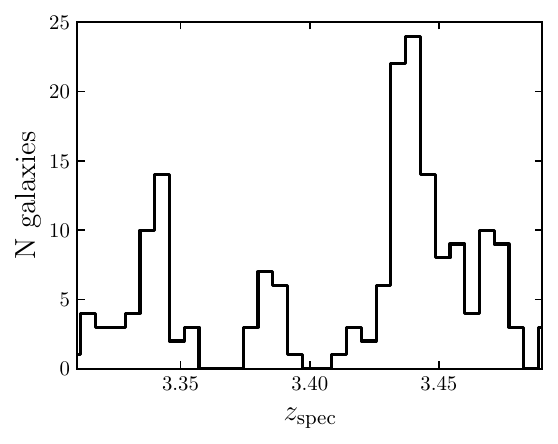}
    \caption{Distribution of spectroscopically confirmed galaxies at $3.3<z_{\rm spec}<3.5$ in the CEERS field.}
    \label{fig:zhist}
\end{figure}

\begin{figure}[!htpb]
    \centering
    \includegraphics[width=\columnwidth]{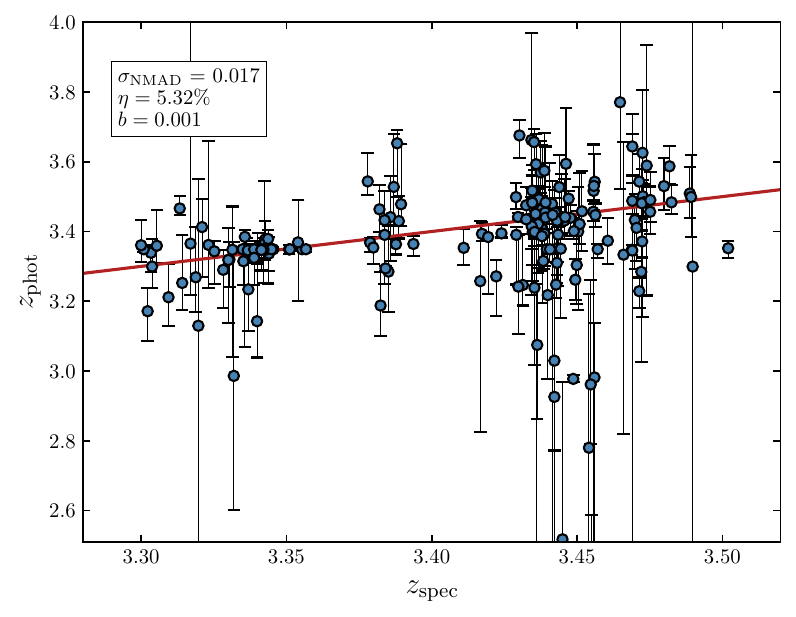}
    \caption{Comparison between spectroscopic redshift and EAZY photometric redshift, for spectroscopically confirmed galaxies in CEERS with $3.3<z_{\rm spec}<3.5$.}
    \label{fig:photz_specz}
\end{figure}

\FloatBarrier
\section{Mass-size relation fits}
\begin{table}[!htbp]
    \centering
    \caption{Fitted and literature mass-size relations.}
    \renewcommand\arraystretch{1.4}
    \setlength{\tabcolsep}{3.2pt}
    \begin{tabular}{c | c | c | c}
        Sample & $\log{(A/{\rm kpc)}}$ & $\alpha$ & Reference\\
        \hline
       Field  & $0.35\pm0.03$ & $0.22\pm0.01$ & This work\\
       Vine SFGs & $0.26\pm0.05$ & $0.16\pm0.03$ & This work\\
       \hline
       $z\sim3.75$ LTGs & $0.41^{+0.04}_{-0.03}$ & $0.25^{+0.05}_{-0.03}$ & \citet{Ward2024_Mass_size_relation}\\
       $z\sim3.75$ QGs & $-0.24^{+0.08}_{-0.09}$ & $0.54^{+0.24}_{-0.28}$ & \citet{Ito2024_QG_mass_size}\\
    \end{tabular}
    \label{tab:mass_size_results}
\end{table}

\clearpage

\end{appendix}
\end{document}